\pdfoutput=1
\documentclass[11pt]{article}

\usepackage{ACL2023}

\usepackage{times}
\usepackage{latexsym}
\usepackage{multirow}
\usepackage{graphicx}
\usepackage{booktabs} 
\usepackage{enumitem}
\usepackage{subfigure}
\usepackage{caption}
\usepackage{amsmath}

\usepackage{xcolor}
\usepackage{pifont} % for checkmark and cross symbols
\usepackage{amssymb} % for checkmark and cross symbols
\definecolor{checkcolor}{rgb}{0,0.6,0} % green color for checkmark
\definecolor{crosscolor}{rgb}{0.9,0,0} % red color for cross
\newcommand{\cmark}{\ding{51}} % checkmark
\newcommand{\xmark}{\ding{55}} % cross

\usepackage[T1]{fontenc}
\usepackage[utf8]{inputenc}

\usepackage{microtype}

\usepackage{inconsolata}

\title{Unleashing the Power of Large Language Models in Zero-shot Relation Extraction via Self-Prompting}

\author{
    Siyi Liu$^{1\dag}$ \hspace{0.5cm} Yang Li$^{2\ast}$ \hspace{0.5cm} Jiang Li$^{2}$ \hspace{0.5cm}
     \textbf{Shan Yang}$^{3}$ \hspace{0.5cm} \textbf{Yunshi Lan}$^{4}$ \\
    $^1$EPFL, Lausanne, Switzerland \\
    $^2$MYBank, Ant Group, Beijing, China \\
    $^3$Beihang University, Beijing, China \\
    $^4$East China Normal University, Beijing, China \\
    \texttt{ssui.liu1022@gmail.com} \hspace{0.5cm} \texttt{ly200170@mybank.cn} \hspace{0.5cm} \texttt{lj311207@mybank.cn} \\
    \texttt{shanyang@buaa.edu.cn} \hspace{0.5cm} \texttt{yslan@dase.ecnu.cn}\\
}

\begin{document}
\maketitle
\begin{abstract}
Recent research in zero-shot Relation Extraction (RE) has focused on using Large Language Models (LLMs) due to their impressive zero-shot capabilities. However, current methods often perform suboptimally, mainly due to a lack of detailed, context-specific prompts needed for understanding various sentences and relations. To address this, we introduce the Self-Prompting framework, a novel method designed to fully harness the embedded RE knowledge within LLMs. Specifically, our framework employs a three-stage diversity approach to prompt LLMs, generating multiple synthetic samples that encapsulate specific relations from scratch. These generated samples act as in-context learning samples, offering explicit and context-specific guidance to efficiently prompt LLMs for RE. Experimental evaluations on benchmark datasets show our approach outperforms existing LLM-based zero-shot RE methods. Additionally, our experiments confirm the effectiveness of our generation pipeline in producing high-quality synthetic data that enhances performance.

\end{abstract}

\section{Introduction}

\begin{figure*}[!ht]
    \centering
    \includegraphics[width=0.8\linewidth]{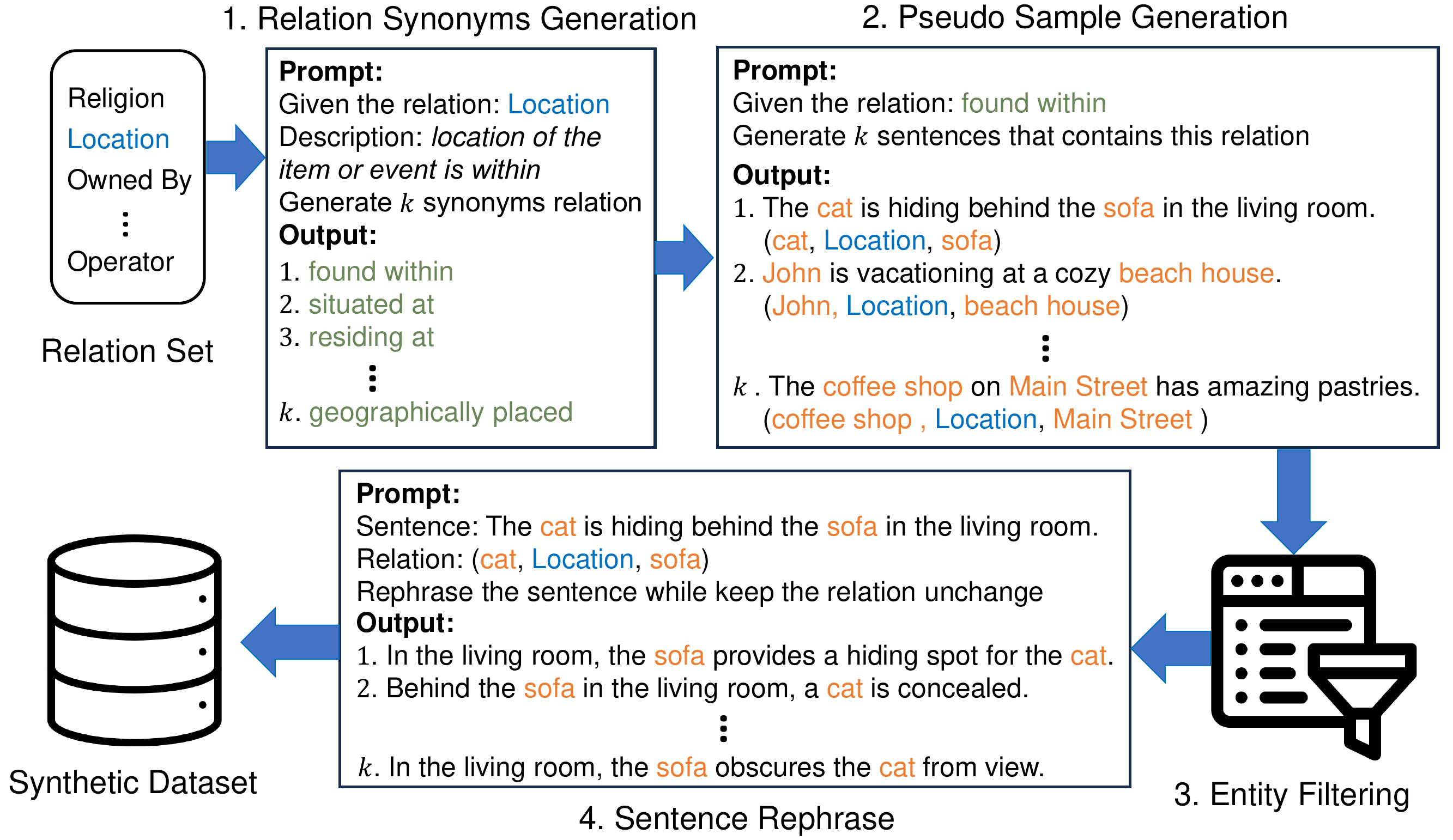}
    \caption{Depiction of the three-stage synthetic sample generation pipeline, where \textcolor{cyan}{blue} indicates candidate relations, \textcolor{lime}{green} signifies synonym relations, and \textcolor{orange}{orange} highlights entities within sentences.}
    \label{fig:pipeline}
\end{figure*}

\renewcommand{\thefootnote}{}
\footnotetext{$^{\dag}$The work is performed when Siyi Liu is an intern in MYBank, Ant Group.}
\footnotetext{$^{\ast}$Corresponding Author.}
\renewcommand{\thefootnote}{\arabic{footnote}}

Recent advances in Large Language Models (LLMs) have significantly progressed Natural Language Processing (NLP). Leveraging LLMs' potential in zero-shot learning, there is growing interest in applying their capabilities to zero-shot Relation Extraction (RE) \cite{han2018fewrel, chen2021zs}, which identifies relationships between entities in text without extensive data annotation. Specifically, current methods convert the RE task into a Question Answering (QA) task by reformulating sentences as questions and candidate relations as options \cite{zhang-etal-2023-aligning}. Further advancements integrate a self-consistency approach \cite{wang2022self} within QA to reduce uncertainty through majority voting \cite{li2023revisiting}. 

However, current methods frequently demonstrate suboptimal performance, mainly because of insufficient guidance for RE. The intricate demands of RE necessitate more detailed and context-specific prompts to effectively comprehend the diverse and complex nature of sentences and relations \cite{bassignanaplank2022mean, zhao2023comprehensive}.

Inspired by recent studies on \textbf{Self-Prompting} \cite{li2022self, wan-etal-2023-better, wan2023universal}—that is, \textit{employing the outputs generated by LLMs themselves as prompts}—our research introduces a novel prompting paradigm for RE. This paradigm leverages LLMs' inherent capabilities to create synthetic RE data tailored to specific relations. When using LLMs for RE from specific sentences, these synthetic samples, enriched with essential relational knowledge, serve as effective in-context demonstrations. 

To be specific, for each distinct relation, we initially prompt LLMs to generate a corresponding sample comprising a sentence and its related relation triple. However, directly prompting LLMs to generate samples may result in a lack of \textbf{diversity} and \textbf{coverage} \cite{chung2023increasing, yu2024wavecoder}, which are crucial for in-context learning \cite{levy2022diverse, liqiu2023finding}. Consequently, to guarantee the quality and comprehensive coverage of these synthetic samples, we implement a three-stage diversification strategy:
\textbf{1. Relation Synonyms}: Utilizing LLMs, we generate synonyms for each relation, broadening semantic understanding and data variability. \textbf{2. Entity Filtering}: We filter out generated samples containing high-frequency entities to prevent repetitions, thereby ensuring the uniqueness of each data point. \textbf{3. Sentence Rephrase}: By rephrasing generated sentences, we introduce structural variation and enhance the linguistic complexity of our dataset. The integration of these diversification methods results in a robust and varied set of synthetic data for RE. During inference, we select salient examples from this synthetic dataset as in-context demonstrations for each test sample, concatenating them with the test question to form the final input sequence for the LLM to generate the final answer.

To verify our method's effectiveness, we evaluated it across multiple zero-shot RE datasets. Compared to previous prompting strategies for LLM-based zero-shot RE SoTA, our method significantly outperforms them. Furthermore, extensive experiments have shown that our three-stage diversification strategy substantially enhances the diversity and coverage of in-context samples, thereby boosting model performance.

\section{Methodology}

\subsection{Relation Synonyms}
\label{sec:syn}
Our methodology's initial phase generates relation synonyms to broaden relation synonym coverage. This strategy recognizes that a dataset's relation often represents a broad concept, covering various synonymous or semantically related terms. As detailed in Figure \ref{fig:pipeline}, Step 1, we utilize LLMs to generate $k$ synonyms for each targeted relation. To ensure the generated synonyms align with the relation's meaning, we provide the description of the relation to the LLMs. We then integrate the original relation with these synonyms to form a comprehensive semantic group. This process ensures the group encompasses the original relation alongside its synonyms, enhancing the relation's contextual comprehension. 

\subsection{Synthetic Sample Generation with Entity Filtering}

After establishing semantic groups for each relation, we then prompt LLMs to create synthetic samples (as shown in Step 2 of Figure \ref{fig:pipeline}). However, these directly generated samples often lack sufficient entity coverage, reflecting the real world's complexity and variability in sentence structures. Such reliance on LLMs may result in a skewed distribution of entities, favoring those frequently found in pretraining and Supervised Fine-Tuning (SFT) data \cite{li-etal-2023-synthetic, xu2023knowledgeinfused}. This issue is not unique to our approach but has also been observed in other LLM-based domain-specific data generation efforts (e.g., \citet{li-etal-2023-synthetic, xu2023knowledgeinfused}). 

To ensure comprehensive entity coverage, we implement a filtration mechanism for generated samples. This method discards samples with entities appearing more than $n$ times, preventing overrepresentation. Conversely, samples with less frequent entities are retained, and their occurrence counts are updated. This strategy mitigates bias towards prevalent entities, promoting a diverse and balanced entity representation in our synthetic sample collection.

\subsection{Sentence Rephrase}

In our Self-Prompting framework, semantic coverage is crucial for sample diversity. The placement of subject and object entities in sentences can vary widely, and relations may be expressed implicitly or explicitly. Thus, incorporating diverse linguistic forms in synthetic data is essential.

To address this, we use LLMs to rephrase each sentence in the synthetic samples, creating $r$ variants with similar meanings (as shown in Figure \ref{fig:pipeline}, Step 4). These rephrased versions differ in structure but maintain the original relation, whether explicit or implicit. This method enhances the range of linguistic expressions in our dataset while ensuring consistent portrayal of the relationship across different semantic representations.

\subsection{Self-Prompting Inference}

In the inference phase for a given test sentence, we retrieve $d$ semantically similar samples as in-context demonstrations. This involves encoding the test sentence with the sentence embedding model and selecting the most similar examples from our sample set using cosine similarity.

To organize the retrieved samples effectively, we implement a ranking strategy based on similarity scores \cite{liu-etal-2022-makes}, arranging samples from the lowest to the highest score. This method positions the most relevant sample nearest to the test sentence, optimizing the impact of contextually appropriate samples on the LLM's inference process. 

\subsection{Addressing the Error Propagation Problem}
Error propagation is a critical concern in complex pipelines like ours, where early inaccuracies can accumulate and adversely impact downstream tasks. For instance, if incorrect or imprecise synonyms are generated during the Relation Synonyms Generation step, these errors may cascade through subsequent stages, resulting in further inaccuracies in relation extraction and other tasks that rely on these synonyms. To mitigate this risk, we incorporate relation descriptions (as detailed in Section \ref{sec:syn} and Table \ref{table:prompt} in the appendix). This enables the language model to better grasp the context of the relations, thereby enhancing the accuracy of synonym generation, as demonstrated in Table \ref{table:performance_comparison}.

\begin{table*}[h]
\centering

\resizebox{0.8\textwidth}{!}{
    \begin{tabular}{l l ccc | ccc | ccc}
    % \hline
    \toprule
    \multirow{2}{*}{Type} & \multirow{2}{*}{Method} & \multicolumn{3}{c}{$m=5$} & \multicolumn{3}{c}{$m=10$} & \multicolumn{3}{c}{$m=15$} \\
    % \cline{3-5} \cline{6-8} \cline{9-11}
    & & Prec. & Rec. & F1 & Prec. & Rec. & F1 & Prec. & Rec. & F1 \\
    % \hline
    \midrule
    \multirow{7}{*}{Zero-shot} & R-BERT & 39.22 & 43.27 & 41.15 & 26.18 & 29.69 & 27.82 & 17.31 & 18.82 & 18.03 \\
    & ESIM & 48.58 & 47.74 & 48.16 & 44.12 & 45.46 & 44.78 & 27.31 & 29.62 & 28.42 \\
    & CIM & 49.63 & 48.81 & 49.22 & 46.54 & 47.90 & 45.57 & 29.17 & 30.58 & 29.86 \\
    & ZS-BERT & 71.54 & 72.39 & 71.96 & 60.51 & 60.98 & 60.74 & 34.12 & 34.38 & 34.25 \\
    & RE-Prompt (NoGen) & 51.78 & 46.76 & 48.93 & 54.87 & 36.52 & 43.80 & 54.45 & 29.43 & 37.45 \\
    & RE-Prompt & 70.66 & \textbf{83.75} & 76.63 & 68.51 & \textbf{74.76} & 71.50 & 63.69 & \textbf{67.93} & 65.74 \\
    & RE-Matching & \textbf{78.19} & \underline{78.41} & \textbf{78.30} & \underline{74.39} & 73.54 & \underline{73.96} & \underline{67.31} & 67.33 & \underline{67.32} \\
    \midrule
    \multirow{3}{*}{LLMs} & Vanilla & 74.45 & 59.25 & 65.98 & 61.15 & 57.68 & 59.36 & 57.82 & 61.27 & 59.01 \\
    & SumAsk & 75.64 & 70.96 & 73.32 & 62.31 & 61.08 & 61.69 & 43.55 & 40.27 & 41.85 \\
    & Self-Prompting & \underline{78.13} & 77.01 & \underline{77.57} & \textbf{75.21} & \underline{74.43} & \textbf{74.81} & \textbf{69.95} & \underline{67.50} & \textbf{68.70} \\
    \bottomrule
    \end{tabular}
}
\caption{Main results on Wiki-ZSL. We mark the best results in \textbf{bold}, the second-best \underline{underlined}. The results of the baselines are retrieved from \citet{li2023revisiting} and \citet{zhao2023re}.}
\label{table-main-wiki}
\end{table*}

\begin{table*}[h]
\centering
\resizebox{0.8\textwidth}{!}{
    \begin{tabular}{l l ccc | ccc | ccc}
    \toprule
    \multirow{2}{*}{Type} & \multirow{2}{*}{Method} & \multicolumn{3}{c}{$m=5$} & \multicolumn{3}{c}{$m=10$} & \multicolumn{3}{c}{$m=15$} \\
    % \cline{3-5} \cline{6-8} \cline{9-11}
    & & Prec. & Rec. & F1 & Prec. & Rec. & F1 & Prec. & Rec. & F1 \\
    \midrule
    \multirow{7}{*}{Zero-shot} & R-BERT & 42.19 & 48.61 & 45.17 & 25.52 & 33.02 & 28.20 & 16.95 & 19.37 & 18.08 \\
    & ESIM & 56.27 & 58.44 & 57.33 & 42.89 & 44.17 & 43.52 & 29.15 & 31.59 & 30.32 \\
    & CIM & 58.05 & 61.92 & 59.92 & 47.39 & 49.11 & 48.23 & 31.83 & 33.06 & 32.43 \\
    & ZS-BERT & 76.96 & 78.86 & 77.90 & 56.92 & 57.59 & 57.25 & 35.54 & 38.19 & 36.82 \\
    & RE-Prompt (NoGen) & 72.36 & 58.61 & 64.57 & 66.47 & 48.28 & 55.61 & 66.49 & 40.05 & 49.38 \\
    & RE-Prompt & 90.15 & 88.50 & 89.30 & 80.33 & 79.62 & 79.96 & \underline{74.33} & 72.51 & 73.40 \\
    & RE-Matching & \textbf{92.82} & \textbf{92.34} & \textbf{92.58} & \textbf{83.21} & \underline{82.64} & \textbf{82.93} & 73.80 & \underline{73.52} & \underline{73.66} \\
    \midrule
    \multirow{3}{*}{LLMs} & Vanilla & 91.70 & 88.87 & 90.26 & 72.64 & 76.12 & 74.34 & 65.46 & 65.50 & 65.48 \\
    & SumAsk & 78.27 & 72.55 & 75.30 & 64.77 & 60.94 & 62.80 & 44.76 & 41.13 & 42.87 \\
    &  Self-Prompting & \underline{90.47} & \underline{89.83} & \underline{90.19} & \underline{81.15} & \textbf{83.02} & \underline{82.07} & \textbf{75.54} & \textbf{78.01} & \textbf{76.76} \\
    \bottomrule
    \end{tabular}
}
\caption{Main results on FewRel. We mark the best results in \textbf{bold}, the second-best \underline{underlined}. The results of the baselines are retrieved from \citet{li2023revisiting} and \citet{zhao2023re}.}
\label{table-main-fewrel}
\end{table*}

\section{Experimental Setup}
\subsection{Datasets}
We evaluate our methods on four RE datasets:
(1) \textbf{FewRel} \cite{han2018fewrel}, (2) \textbf{Wiki-ZSL} \cite{sorokin2017context}, (3) \textbf{TACRED.} \cite{zhang2017position}, (4) \textbf{SemEval} \cite{hendrickx2009semeval}. Further details about the dataset preprocessing and data statistics are in Appendix \ref{sec:data}.

\subsection{Implementation Details}
In our study, we employed ChatGPT with the API version \texttt{gpt-3.5-turbo-0301}, in line with previous research \cite{zhang-etal-2023-aligning, li2023revisiting}. The text embedding model utilized was \texttt{text-embedding-ada-002}, accessed via the OpenAI API. For more details about hyperparameter setting on API usage and synthetic sample generation, please refer to Appendix \ref{hp-setting}.

\subsection{Baselines}
\textbf{Zero-shot Baselines}: 
For the FewRel and WikiZSL datasets, our baseline models include R-BERT \cite{wu2019enriching}, ESIM \cite{chen2017enhanced}, CIM \cite{rocktaschel2016reasoning}, ZS-BERT \cite{chen2021zs}, RE-Prompt \cite{chia2022relationprompt} and RE-Matching \cite{zhao2023re}. For RE-Prompt, the NoGen variant represents outcomes without data generation. Regarding TACRED and SemEval, our baseline comparisons involve NLI \cite{sainz2021label} and SuRE \cite{lu-etal-2022-summarization}. Here, the underlying base models are DeBERTa-XLarge \cite{he2020deberta} for NLI and PEGASUS-Large \cite{zhang2020pegasus} for SuRE. \\
\textbf{LLMs Baselines}: 
In evaluating prompt-based LLM baselines, we selected SumAsk \cite{li2023revisiting} and QA4RE \cite{zhang-etal-2023-aligning} for comparison. We also present the performance using a vanilla prompt strategy (denoted as \textbf{Vanilla}). This approach involves directly prompting LLMs to deduce the relation within a sentence, absent any in-context demonstrations ($d=0$).

\section{Results and Analysis}

\subsection{Main Results}

Our evaluation of zero-shot prompting in LLMs, conducted on the FewRel and Wiki-ZSL datasets (as detailed in Tables \ref{table-main-wiki} and \ref{table-main-fewrel}), shows competitive performance against existing zero-shot RE methods. Notably, our Self-Prompting technique significantly enhances ChatGPT's performance over Vanilla prompting, outperforming the RE-Prompt method in most scenarios and markedly surpassing the SumAsk prompt strategy.

As the number of unseen relations ($m$) increases, predicting the correct relation becomes more challenging due to a broader range of choices. However, the benefits of Self-Prompting become more apparent, while Vanilla and SumAsk approaches show significant performance declines. This is likely because in-context demonstrations effectively narrow down potential relations. As a result, Self-Prompting better guides LLMs in inferring correct relations and demonstrates greater resilience with increasing relations.

Further validation is demonstrated through the application of our method on the TACRED and SemEval datasets. As shown in Table \ref{table-main-tacred}, our Self-Prompting technique achieved the highest F1 score on the SemEval dataset and the second-highest on TACRED, outperforming other zero-shot methods and significantly exceeding the performance of the QA4RE prompt strategy. This highlights the effectiveness of our approach, particularly given QA4RE’s established performance.

\begin{table}% [h]
\centering
\resizebox{0.40\textwidth}{!}{
    \begin{tabular}{lcccccc}
    \hline Datasets & \multicolumn{3}{c}{ TACRED } & \multicolumn{3}{c}{ SemEval } \\
    % \cline{2-4} \cline{5-7} \\
    & Prec. & Rec. & F1 & Prec. & Rec. & F1 \\
    \hline 
    NLI$_{\mathrm{DeBERTa}}$ & \underline{42.9} & \textbf{76.9} & \underline{55.1} & 22.0 & 25.7 & 23.7 \\
    SuRE$_{\mathrm{PEGASUS}}$ & 13.8 & 51.7 & 21.8 & 0.0 & 0.0 & 0.0 \\
    Vanilla & 32.1 & \underline{74.8} & 44.9 & 18.2 & 20.8 & 19.4 \\
    SumAsk	& \textbf{62.2}	& 53.8 & \textbf{57.7} & - & - & - \\
    QA4RE  & 32.8 & 68.0 & 44.2 & \underline{29.9} & \underline{35.2} & \underline{32.3} \\
    \midrule
    Self-Prompting    & \underline{56.8} & 57.5 & \underline{57.1} & \textbf{55.3} & \textbf{50.9} & \textbf{52.7} \\
    \hline
    \end{tabular}
}
\caption{Main results on TACRED and SemEval. We mark the best results in \textbf{bold}, the second-best \underline{underlined}. The results of the baselines are retrieved from \cite{zhang-etal-2023-aligning}}
\label{table-main-tacred}
\end{table}

\subsection{Ablation Study on Different Diversity Strategies}

In our ablation study, we systematically examine the impact of different components of our synthetic data generation method on the FewRel and Wiki-ZSL datasets. The absence of each component is denoted by a specific condition in our experiments: \textbf{w/o Rephrasing} (omission of sentence rephrasing), \textbf{w/o Synonyms} (exclusion of relation synonyms generation), \textbf{w/o Entity Filtering} (absence of entity frequency filtering), \textbf{w/o All} (direct generation without any enhancements), \textbf{Vanilla} (zero-shot learning without any generated samples, serving as a baseline), and \textbf{Complete} (all diversification strategies are included).

As we can see in Table \ref{table:performance_comparison}, the findings emphasize the importance of each component. Removing sentence rephrasing slightly decreases Precision and F1 scores. Excluding relation synonym generation results in a more substantial drop across all metrics, highlighting the importance of synonyms for capturing semantic breadth. Omitting entity frequency filtering significantly impacts Recall, indicating that entity variety is crucial for comprehensive relation extraction.

\begin{table}% [h]
\centering
\resizebox{0.45\textwidth}{!}{
    \begin{tabular}{lcccccc}
    \hline 
    Strategy & \multicolumn{3}{c}{Wiki-ZSL} & \multicolumn{3}{c}{FewRel} \\ 
     & Prec. & Rec. & F1 & Prec. & Rec. & F1 \\ 
    \hline 
    Vanilla & 70.1 & 72.4 & 71.2 & 81.8 & 78.2 & 80.1 \\ 
    w/o Synonyms & 73.0 & 73.3 & 73.2 & 82.1 & 80.3 & 81.3 \\ 
    w/o Entity Filtering & 74.9 & 75.3 & 75.1 & 82.7 & 80.8 & 81.7 \\ 
    w/o Rephrase & 77.0 & 75.7 & 76.3 & 84.5 & 81.9 & 83.2 \\ 
    w/o All & 65.7 & 67.9 & 66.8 & 78.5 & 81.4 & 80.0 \\ 
    Complete & \textbf{82.4} & \textbf{77.7} & \textbf{80.0} & \textbf{85.4} & \textbf{83.3} & \textbf{84.3} \\ 
    \hline 
    \end{tabular}
}
\caption{Performance comparison of different strategies on FewRel and Wiki-ZSL datasets ($m=10$).}
\label{table:performance_comparison}
\end{table}

Moreover, directly prompting LLMs to generate samples and using them for inference impairs the model's performance, as evidenced by the w/o All condition, which underperforms compared to the Vanilla baseline. This suggests that unrefined sample generation can adversely affect the quality of RE. In contrast, our method (Complete), which incorporates all techniques, consistently outperforms the other conditions. It notably secures the highest Precision, Recall, and F1 scores across both datasets, confirming our comprehensive approach's effectiveness.

\section{Conclusion}

In this study, we introduced the Self-Prompting framework to optimize zero-shot RE in LLMs. Our three-stage diversification strategy generates synthetic samples, enhancing LLMs' accuracy and efficiency in RE. Experimental results on benchmark datasets demonstrate our method's effectiveness, surpassing existing LLM-based zero-shot RE techniques. Further experiments confirm that our strategy successfully addresses the challenges of diversity and coverage in synthetic sample generation, thereby improving model performance.

\section*{Limitations}
While our Self-Prompting method demonstrates promising outcomes in zero-shot RE, it also presents certain limitations. Firstly, the selection of appropriate in-context demonstrations from synthetic datasets requires further exploration, as improper samples may introduce noise, adversely affecting LLM performance in zero-shot RE. Additionally, the performance of our Self-Prompting method on domain-specific data remains uncertain, given that domain-specific data generation poses an ongoing challenge. We acknowledge these issues and leave them for future work to address.

\section*{Ethics Statement}
This work employs text generated by Large Language Models (LLMs), which may inadvertently produce content with ethical or safety concerns. However, given that ChatGPT, the LLM utilized in our experiments, is rigorously designed to minimize the generation of untrustworthy and harmful information, and considering the specific context of zero-shot relation extraction, we contend that the ethical considerations related to this research are limited. Consequently, a detailed discussion of these issues is deemed unnecessary.

\section*{Acknowledgements}
This work was supported by Ant Group Research Intern Program.

\bibliography{anthology,custom}

\begin{thebibliography}{40}
\expandafter\ifx\csname natexlab\endcsname\relax\def\natexlab#1{#1}\fi

\bibitem[{Bai et~al.(2023)Bai, Bai, Chu, Cui, Dang, Deng, Fan, Ge, Han, Huang, Hui, Ji, Li, Lin, Lin, Liu, Liu, Lu, Lu, Ma, Men, Ren, Ren, Tan, Tan, Tu, Wang, Wang, Wang, Wu, Xu, Xu, Yang, Yang, Yang, Yang, Yao, Yu, Yuan, Yuan, Zhang, Zhang, Zhang, Zhang, Zhou, Zhou, Zhou, and Zhu}]{bai2023qwen}
Jinze Bai, Shuai Bai, Yunfei Chu, Zeyu Cui, Kai Dang, Xiaodong Deng, Yang Fan, Wenbin Ge, Yu~Han, Fei Huang, Binyuan Hui, Luo Ji, Mei Li, Junyang Lin, Runji Lin, Dayiheng Liu, Gao Liu, Chengqiang Lu, Keming Lu, Jianxin Ma, Rui Men, Xingzhang Ren, Xuancheng Ren, Chuanqi Tan, Sinan Tan, Jianhong Tu, Peng Wang, Shijie Wang, Wei Wang, Shengguang Wu, Benfeng Xu, Jin Xu, An~Yang, Hao Yang, Jian Yang, Shusheng Yang, Yang Yao, Bowen Yu, Hongyi Yuan, Zheng Yuan, Jianwei Zhang, Xingxuan Zhang, Yichang Zhang, Zhenru Zhang, Chang Zhou, Jingren Zhou, Xiaohuan Zhou, and Tianhang Zhu. 2023.
\newblock \href {http://arxiv.org/abs/2309.16609} {Qwen technical report}.

\bibitem[{Bassignana and Plank(2022)}]{bassignanaplank2022mean}
Elisa Bassignana and Barbara Plank. 2022.
\newblock \href {https://doi.org/10.18653/v1/2022.acl-srw.7} {What do you mean by relation extraction? a survey on datasets and study on scientific relation classification}.
\newblock In \emph{Proceedings of the 60th Annual Meeting of the Association for Computational Linguistics: Student Research Workshop}, pages 67--83, Dublin, Ireland. Association for Computational Linguistics.

\bibitem[{Chen and Li(2021)}]{chen2021zs}
Chih-Yao Chen and Cheng-Te Li. 2021.
\newblock Zs-bert: Towards zero-shot relation extraction with attribute representation learning.
\newblock In \emph{Proceedings of the 2021 Conference of the North American Chapter of the Association for Computational Linguistics: Human Language Technologies}, pages 3470--3479.

\bibitem[{Chen et~al.(2017)Chen, Zhu, Ling, Wei, Jiang, and Inkpen}]{chen2017enhanced}
Qian Chen, Xiaodan Zhu, Zhen-Hua Ling, Si~Wei, Hui Jiang, and Diana Inkpen. 2017.
\newblock Enhanced lstm for natural language inference.
\newblock In \emph{Proceedings of the 55th Annual Meeting of the Association for Computational Linguistics (Volume 1: Long Papers)}, pages 1657--1668.

\bibitem[{Chia et~al.(2022)Chia, Bing, Poria, and Si}]{chia2022relationprompt}
Yew~Ken Chia, Lidong Bing, Soujanya Poria, and Luo Si. 2022.
\newblock Relationprompt: Leveraging prompts to generate synthetic data for zero-shot relation triplet extraction.
\newblock In \emph{Findings of the Association for Computational Linguistics: ACL 2022}, pages 45--57.

\bibitem[{Chung et~al.(2023)Chung, Kamar, and Amershi}]{chung2023increasing}
John Chung, Ece Kamar, and Saleema Amershi. 2023.
\newblock \href {https://doi.org/10.18653/v1/2023.acl-long.34} {Increasing diversity while maintaining accuracy: Text data generation with large language models and human interventions}.
\newblock In \emph{Proceedings of the 61st Annual Meeting of the Association for Computational Linguistics (Volume 1: Long Papers)}, pages 575--593, Toronto, Canada. Association for Computational Linguistics.

\bibitem[{Han et~al.(2018)Han, Zhu, Yu, Wang, Yao, Liu, and Sun}]{han2018fewrel}
Xu~Han, Hao Zhu, Pengfei Yu, Ziyun Wang, Yuan Yao, Zhiyuan Liu, and Maosong Sun. 2018.
\newblock Fewrel: A large-scale supervised few-shot relation classification dataset with state-of-the-art evaluation.
\newblock In \emph{Proceedings of the 2018 Conference on Empirical Methods in Natural Language Processing}. Association for Computational Linguistics.

\bibitem[{He et~al.(2020)He, Liu, Gao, and Chen}]{he2020deberta}
Pengcheng He, Xiaodong Liu, Jianfeng Gao, and Weizhu Chen. 2020.
\newblock Deberta: Decoding-enhanced bert with disentangled attention.
\newblock In \emph{International Conference on Learning Representations}.

\bibitem[{Hendrickx et~al.(2009)Hendrickx, Kim, Kozareva, Nakov, {\'O}~S{\'e}aghdha, Pad{\'o}, Pennacchiotti, Romano, and Szpakowicz}]{hendrickx2009semeval}
Iris Hendrickx, Su~Nam Kim, Zornitsa Kozareva, Preslav Nakov, Diarmuid {\'O}~S{\'e}aghdha, Sebastian Pad{\'o}, Marco Pennacchiotti, Lorenza Romano, and Stan Szpakowicz. 2009.
\newblock \href {https://aclanthology.org/W09-2415} {{S}em{E}val-2010 task 8: Multi-way classification of semantic relations between pairs of nominals}.
\newblock In \emph{Proceedings of the Workshop on Semantic Evaluations: Recent Achievements and Future Directions ({SEW}-2009)}, pages 94--99, Boulder, Colorado. Association for Computational Linguistics.

\bibitem[{Kojima et~al.(2022)Kojima, Gu, Reid, Matsuo, and Iwasawa}]{kojima2022large}
Takeshi Kojima, Shixiang~Shane Gu, Machel Reid, Yutaka Matsuo, and Yusuke Iwasawa. 2022.
\newblock Large language models are zero-shot reasoners.
\newblock \emph{Advances in neural information processing systems}, 35:22199--22213.

\bibitem[{Levy et~al.(2022)Levy, Bogin, and Berant}]{levy2022diverse}
Itay Levy, Ben Bogin, and Jonathan Berant. 2022.
\newblock Diverse demonstrations improve in-context compositional generalization.
\newblock \emph{arXiv preprint arXiv:2212.06800}.

\bibitem[{Levy et~al.(2017)Levy, Seo, Choi, and Zettlemoyer}]{levy2017zero}
Omer Levy, Minjoon Seo, Eunsol Choi, and Luke Zettlemoyer. 2017.
\newblock Zero-shot relation extraction via reading comprehension.
\newblock In \emph{Proceedings of the 21st Conference on Computational Natural Language Learning (CoNLL 2017)}, pages 333--342.

\bibitem[{Li et~al.(2023{\natexlab{a}})Li, Wang, and Ke}]{li2023revisiting}
Guozheng Li, Peng Wang, and Wenjun Ke. 2023{\natexlab{a}}.
\newblock Revisiting large language models as zero-shot relation extractors.
\newblock \emph{arXiv preprint arXiv:2310.05028}.

\bibitem[{Li et~al.(2022)Li, Zhang, and Zhao}]{li2022self}
Junlong Li, Zhuosheng Zhang, and Hai Zhao. 2022.
\newblock Self-prompting large language models for open-domain qa.
\newblock \emph{arXiv preprint arXiv:2212.08635}.

\bibitem[{Li and Qiu(2023)}]{liqiu2023finding}
Xiaonan Li and Xipeng Qiu. 2023.
\newblock \href {https://doi.org/10.18653/v1/2023.findings-emnlp.411} {Finding support examples for in-context learning}.
\newblock In \emph{Findings of the Association for Computational Linguistics: EMNLP 2023}, pages 6219--6235, Singapore. Association for Computational Linguistics.

\bibitem[{Li et~al.(2023{\natexlab{b}})Li, Zhu, Lu, and Yin}]{li-etal-2023-synthetic}
Zhuoyan Li, Hangxiao Zhu, Zhuoran Lu, and Ming Yin. 2023{\natexlab{b}}.
\newblock \href {https://doi.org/10.18653/v1/2023.emnlp-main.647} {Synthetic data generation with large language models for text classification: Potential and limitations}.
\newblock In \emph{Proceedings of the 2023 Conference on Empirical Methods in Natural Language Processing}, pages 10443--10461, Singapore. Association for Computational Linguistics.

\bibitem[{Liu et~al.(2022{\natexlab{a}})Liu, Shen, Zhang, Dolan, Carin, and Chen}]{liu-etal-2022-makes}
Jiachang Liu, Dinghan Shen, Yizhe Zhang, Bill Dolan, Lawrence Carin, and Weizhu Chen. 2022{\natexlab{a}}.
\newblock \href {https://doi.org/10.18653/v1/2022.deelio-1.10} {What makes good in-context examples for {GPT}-3?}
\newblock In \emph{Proceedings of Deep Learning Inside Out (DeeLIO 2022): The 3rd Workshop on Knowledge Extraction and Integration for Deep Learning Architectures}, pages 100--114, Dublin, Ireland and Online. Association for Computational Linguistics.

\bibitem[{Liu et~al.(2022{\natexlab{b}})Liu, Liu, Lu, Welleck, West, Le~Bras, Choi, and Hajishirzi}]{liu2022generated}
Jiacheng Liu, Alisa Liu, Ximing Lu, Sean Welleck, Peter West, Ronan Le~Bras, Yejin Choi, and Hannaneh Hajishirzi. 2022{\natexlab{b}}.
\newblock Generated knowledge prompting for commonsense reasoning.
\newblock In \emph{Proceedings of the 60th Annual Meeting of the Association for Computational Linguistics (Volume 1: Long Papers)}, pages 3154--3169.

\bibitem[{Lu et~al.(2022)Lu, Hsu, Zhou, Ma, and Chen}]{lu-etal-2022-summarization}
Keming Lu, I-Hung Hsu, Wenxuan Zhou, Mingyu~Derek Ma, and Muhao Chen. 2022.
\newblock \href {https://doi.org/10.18653/v1/2022.findings-emnlp.490} {Summarization as indirect supervision for relation extraction}.
\newblock In \emph{Findings of the Association for Computational Linguistics: EMNLP 2022}, pages 6575--6594, Abu Dhabi, United Arab Emirates. Association for Computational Linguistics.

\bibitem[{Neubig and He(2023)}]{Neubig_Zeno_GPT_Machine_2023}
Graham Neubig and Zhiwei He. 2023.
\newblock {Zeno GPT Machine Translation Report}.

\bibitem[{Rocktaschel et~al.(2016)Rocktaschel, Grefenstette, Hermann, Kocisky, and Blunsom}]{rocktaschel2016reasoning}
Tim Rocktaschel, Edward Grefenstette, Karl~Moritz Hermann, Tomas Kocisky, and Phil Blunsom. 2016.
\newblock Reasoning about entailment with neural attention.
\newblock In \emph{International Conference on Learning Representations (ICLR)}.

\bibitem[{Sainz et~al.(2021)Sainz, de~Lacalle, Labaka, Barrena, and Agirre}]{sainz2021label}
Oscar Sainz, Oier~Lopez de~Lacalle, Gorka Labaka, Ander Barrena, and Eneko Agirre. 2021.
\newblock Label verbalization and entailment for effective zero and few-shot relation extraction.
\newblock In \emph{Proceedings of the 2021 Conference on Empirical Methods in Natural Language Processing}, pages 1199--1212.

\bibitem[{Sorokin and Gurevych(2017)}]{sorokin2017context}
Daniil Sorokin and Iryna Gurevych. 2017.
\newblock Context-aware representations for knowledge base relation extraction.
\newblock In \emph{Proceedings of the 2017 Conference on Empirical Methods in Natural Language Processing}, pages 1784--1789.

\bibitem[{Wan et~al.(2023{\natexlab{a}})Wan, Sun, Dai, Arik, and Pfister}]{wan-etal-2023-better}
Xingchen Wan, Ruoxi Sun, Hanjun Dai, Sercan Arik, and Tomas Pfister. 2023{\natexlab{a}}.
\newblock \href {https://doi.org/10.18653/v1/2023.findings-acl.216} {Better zero-shot reasoning with self-adaptive prompting}.
\newblock In \emph{Findings of the Association for Computational Linguistics: ACL 2023}, pages 3493--3514, Toronto, Canada. Association for Computational Linguistics.

\bibitem[{Wan et~al.(2023{\natexlab{b}})Wan, Sun, Nakhost, Dai, Eisenschlos, Arik, and Pfister}]{wan2023universal}
Xingchen Wan, Ruoxi Sun, Hootan Nakhost, Hanjun Dai, Julian~Martin Eisenschlos, Sercan~O Arik, and Tomas Pfister. 2023{\natexlab{b}}.
\newblock Universal self-adaptive prompting.
\newblock \emph{arXiv preprint arXiv:2305.14926}.

\bibitem[{Wang et~al.(2022{\natexlab{a}})Wang, Srikumar, Hajishirzi, and Smith}]{wang2022elaboration}
Wenya Wang, Vivek Srikumar, Hanna Hajishirzi, and Noah~A Smith. 2022{\natexlab{a}}.
\newblock Elaboration-generating commonsense question answering at scale.
\newblock \emph{arXiv preprint arXiv:2209.01232}.

\bibitem[{Wang et~al.(2022{\natexlab{b}})Wang, Wei, Schuurmans, Le, Chi, Narang, Chowdhery, and Zhou}]{wang2022self}
Xuezhi Wang, Jason Wei, Dale Schuurmans, Quoc~V Le, Ed~H Chi, Sharan Narang, Aakanksha Chowdhery, and Denny Zhou. 2022{\natexlab{b}}.
\newblock Self-consistency improves chain of thought reasoning in language models.
\newblock In \emph{The Eleventh International Conference on Learning Representations}.

\bibitem[{Wei et~al.(2022)Wei, Wang, Schuurmans, Bosma, Xia, Chi, Le, Zhou et~al.}]{wei2022chain}
Jason Wei, Xuezhi Wang, Dale Schuurmans, Maarten Bosma, Fei Xia, Ed~Chi, Quoc~V Le, Denny Zhou, et~al. 2022.
\newblock Chain-of-thought prompting elicits reasoning in large language models.
\newblock \emph{Advances in Neural Information Processing Systems}, 35:24824--24837.

\bibitem[{Wei et~al.(2023)Wei, Cui, Cheng, Wang, Zhang, Huang, Xie, Xu, Chen, Zhang et~al.}]{wei2023zero}
Xiang Wei, Xingyu Cui, Ning Cheng, Xiaobin Wang, Xin Zhang, Shen Huang, Pengjun Xie, Jinan Xu, Yufeng Chen, Meishan Zhang, et~al. 2023.
\newblock Zero-shot information extraction via chatting with chatgpt.
\newblock \emph{arXiv preprint arXiv:2302.10205}.

\bibitem[{Wu and He(2019)}]{wu2019enriching}
Shanchan Wu and Yifan He. 2019.
\newblock Enriching pre-trained language model with entity information for relation classification.
\newblock In \emph{Proceedings of the 28th ACM international conference on information and knowledge management}, pages 2361--2364.

\bibitem[{Xu et~al.(2023)Xu, Cui, Yu, Kan, Shi, Zhuang, Jin, Ho, and Yang}]{xu2023knowledgeinfused}
Ran Xu, Hejie Cui, Yue Yu, Xuan Kan, Wenqi Shi, Yuchen Zhuang, Wei Jin, Joyce Ho, and Carl Yang. 2023.
\newblock \href {http://arxiv.org/abs/2311.00287} {Knowledge-infused prompting: Assessing and advancing clinical text data generation with large language models}.

\bibitem[{Ye et~al.(2022)Ye, Gao, Li, Xu, Feng, Wu, Yu, and Kong}]{ye2022zerogen}
Jiacheng Ye, Jiahui Gao, Qintong Li, Hang Xu, Jiangtao Feng, Zhiyong Wu, Tao Yu, and Lingpeng Kong. 2022.
\newblock Zerogen: Efficient zero-shot learning via dataset generation.
\newblock In \emph{Proceedings of the 2022 Conference on Empirical Methods in Natural Language Processing}, pages 11653--11669.

\bibitem[{Yu et~al.(2022)Yu, Iter, Wang, Xu, Ju, Sanyal, Zhu, Zeng, and Jiang}]{yu2022generate}
Wenhao Yu, Dan Iter, Shuohang Wang, Yichong Xu, Mingxuan Ju, Soumya Sanyal, Chenguang Zhu, Michael Zeng, and Meng Jiang. 2022.
\newblock Generate rather than retrieve: Large language models are strong context generators.
\newblock In \emph{The Eleventh International Conference on Learning Representations}.

\bibitem[{Yu et~al.(2024)Yu, Zhang, Shang, Huang, Xu, Zhao, Hu, and Yin}]{yu2024wavecoder}
Zhaojian Yu, Xin Zhang, Ning Shang, Yangyu Huang, Can Xu, Yishujie Zhao, Wenxiang Hu, and Qiufeng Yin. 2024.
\newblock \href {http://arxiv.org/abs/2312.14187} {Wavecoder: Widespread and versatile enhanced instruction tuning with refined data generation}.

\bibitem[{Zhang et~al.(2020)Zhang, Zhao, Saleh, and Liu}]{zhang2020pegasus}
Jingqing Zhang, Yao Zhao, Mohammad Saleh, and Peter Liu. 2020.
\newblock Pegasus: Pre-training with extracted gap-sentences for abstractive summarization.
\newblock In \emph{International Conference on Machine Learning}, pages 11328--11339. PMLR.

\bibitem[{Zhang et~al.(2023{\natexlab{a}})Zhang, Lan, and He}]{zhang2023contrastive}
Junlei Zhang, Zhenzhong Lan, and Junxian He. 2023{\natexlab{a}}.
\newblock Contrastive learning of sentence embeddings from scratch.
\newblock \emph{arXiv preprint arXiv:2305.15077}.

\bibitem[{Zhang et~al.(2023{\natexlab{b}})Zhang, Jimenez~Gutierrez, and Su}]{zhang-etal-2023-aligning}
Kai Zhang, Bernal Jimenez~Gutierrez, and Yu~Su. 2023{\natexlab{b}}.
\newblock \href {https://doi.org/10.18653/v1/2023.findings-acl.50} {Aligning instruction tasks unlocks large language models as zero-shot relation extractors}.
\newblock In \emph{Findings of the Association for Computational Linguistics: ACL 2023}, pages 794--812, Toronto, Canada. Association for Computational Linguistics.

\bibitem[{Zhang et~al.(2017)Zhang, Zhong, Chen, Angeli, and Manning}]{zhang2017position}
Yuhao Zhang, Victor Zhong, Danqi Chen, Gabor Angeli, and Christopher~D Manning. 2017.
\newblock Position-aware attention and supervised data improve slot filling.
\newblock In \emph{Conference on Empirical Methods in Natural Language Processing}.

\bibitem[{Zhao et~al.(2023{\natexlab{a}})Zhao, Zhan, Zhao, Zhang, Gui, Wei, Wang, Peng, and Sun}]{zhao2023re}
Jun Zhao, Wenyu Zhan, Wayne~Xin Zhao, Qi~Zhang, Tao Gui, Zhongyu Wei, Junzhe Wang, Minlong Peng, and Mingming Sun. 2023{\natexlab{a}}.
\newblock Re-matching: A fine-grained semantic matching method for zero-shot relation extraction.
\newblock In \emph{Proceedings of the 61st Annual Meeting of the Association for Computational Linguistics (Volume 1: Long Papers)}, pages 6680--6691.

\bibitem[{Zhao et~al.(2023{\natexlab{b}})Zhao, Deng, Yang, Wang, Zhang, Cheng, Lam, Shen, and Xu}]{zhao2023comprehensive}
Xiaoyan Zhao, Yang Deng, Min Yang, Lingzhi Wang, Rui Zhang, Hong Cheng, Wai Lam, Ying Shen, and Ruifeng Xu. 2023{\natexlab{b}}.
\newblock \href {http://arxiv.org/abs/2306.02051} {A comprehensive survey on deep learning for relation extraction: Recent advances and new frontiers}.

\end{thebibliography}
\bibliographystyle{acl_natbib}

\appendix

\newpage

\section{Related Works}
\subsection{Zero-shot Relation Extraction}

Zero-shot RE has recently become a crucial focus in advancing predictive model capabilities. \citet{levy2017zero} pioneered zero-shot RE, developing models capable of identifying novel relations beyond predefined types. Furthering this field, \citet{sainz2021label} explored the use of smaller Language Models (LMs) fine-tuned on Natural Language Inference (NLI) datasets. Their approach employs an entity-filled relation template matching the test sentence, utilizing inference for relation prediction. \citet{chen2021zs} incorporate text descriptions of both seen and unseen relations. It employs nearest neighbor search for predicting unseen relations, using embeddings of these relations and new sentences. \citet{lu-etal-2022-summarization} framed RE as a summarization task, applying generative models to concisely express the relationships between target entities. However, a persistent challenge with existing zero-shot methods is their heavy reliance on extensive labeled data. Our research focuses on conducting zero-shot RE without any labeled data.

\subsection{LLMs for Zero-shot Relation Extraction}
In the exploration of Zero-shot RE using LLMs, most existing research has concentrated on designing effective prompts to enhance LLMs' extraction performance. For instance, ChatIE \cite{wei2023zero} employs ChatGPT for zero-shot RE, utilizing a two-stage prompting strategy to refine the LLMs' search scope. QA4RE \cite{zhang-etal-2023-aligning} adopts a multiple-choice question-answering format, representing relations through manually crafted templates and assigning LLMs the task of predicting a single character. In a different approach, SumAsk \cite{li2023revisiting} deconstructs the LLMs' reasoning into three distinct stages, thereby aiding them in understanding and interpreting the relationships between subjects and objects. This method is further enriched by the use of self-consistency \cite{wang2022self} to reduce response uncertainty. However, these methods do not fully harness the LLMs' inherent RE capabilities, primarily because of insufficient context-specific prompting. Our work aims to explore the LLMs' RE potential by utilizing Self-Prompting, which focuses on generating context-specific prompts from synthetic samples.

\subsection{Synthetic Data Generation via LLMs}
Recent research has been focused on leveraging the content generated by LLMs to enhance the training of smaller models in various domains. For instance, \citet{ye2022zerogen} applied this technique in classification tasks, \citet{wang2022elaboration} in commonsense question-answering, \citet{zhang2023contrastive} in contrastive learning, and \citet{chia2022relationprompt} in RE. Additionally, another strand of research directly utilizes the outputs from LLMs. Some studies have employed LLMs to generate relevant contexts or background documents as supplementary inputs for QA tasks \cite{yu2022generate, liu2022generated, li2022self}. Others have focused on eliciting detailed reasoning steps, termed chain-of-thought, particularly for solving arithmetic problems \cite{wei2022chain, wan-etal-2023-better, wan2023universal}. In this work, we capitalize on synthetic RE samples generated by LLMs to bolster their capabilities in RE, exploring a novel approach to enhance the effectiveness of these models in this specific task.

\section{Datasets information}
\label{sec:data}

The statistics of the datasets are shown in Table \ref{table:stat-1} and Table \ref{table:stat-2}. Following previous works \cite{zhang-etal-2023-aligning, li2023revisiting}, for the FewRel and Wiki-ZSL datasets, we randomly selected 5 relations for validation and selected a varying number of unseen relations ($m$) for testing, where $m$ could be 5, 10, or 15. To ascertain the robustness of our results, this classification process was repeated five times, and we report the average macro-F1 scores from these iterations. For TACRED and SemEval, we conduct experiments using only the test samples and present the micro-averaged F1 scores. All relations are included except for \textit{none-of-the-above}.

To effectively manage OpenAI API usage and associated costs, we randomly selected 1,000 samples from the test set of each dataset. We ensured that these samples proportionally represented each relation class.

\begin{table}[h]
    \centering
    \resizebox{0.48\textwidth}{!}{
        \begin{tabular}{lccc}
        \toprule
        Dataset & \# samples & \# entities & \# relations \\
        \midrule        
        FewRel & 56,000 & 72,954 & 80 \\
        Wiki-ZSL & 94,383 & 77,623 & 113 \\
        \bottomrule
    \end{tabular}}
\caption{Statistics of FewRel and Wiki-ZSL}
\label{table:stat-1}
\end{table}

\begin{table}[h]
    \centering
    \resizebox{0.49\textwidth}{!}{
        \begin{tabular}{lcccc}
        \toprule
        Dataset & \# train & \# dev & \# test & \# relations \\
        \midrule
        TACRED & 68,124 & 22,631 & 15,509 & 42 \\
        SemEval & 6,507 & 1,493 & 2,717 & 9 \\
        \bottomrule
    \end{tabular}}
\caption{Statistics of TACRED and SemEval }
\label{table:stat-2}
\end{table}

\begin{figure*}[ht]
    \centering
    % Left graph
    \begin{minipage}{.555\textwidth}
        \centering
        \includegraphics[width=0.95\linewidth]{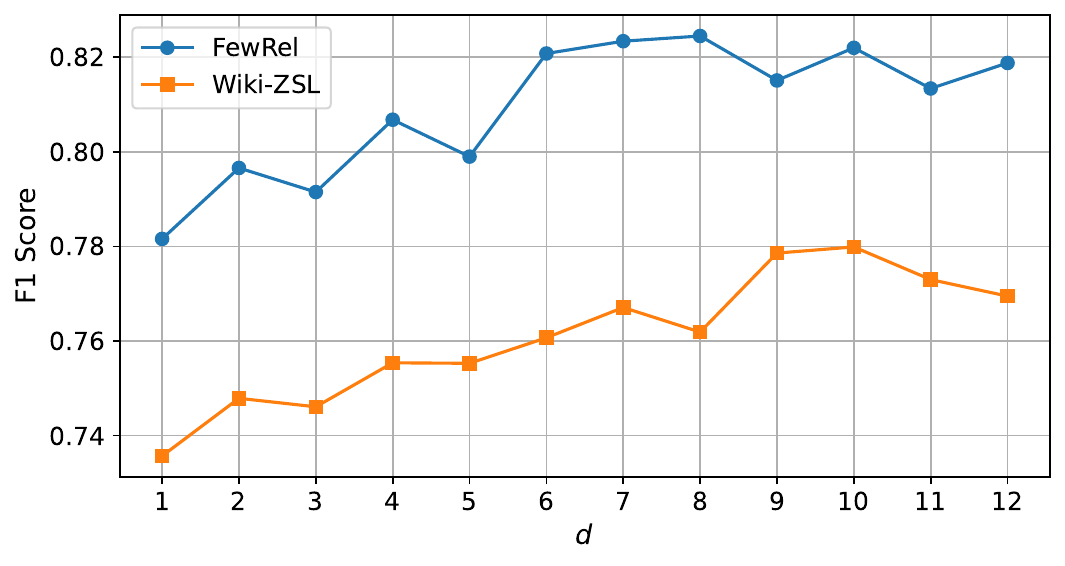}

        \caption{Average F1 when using different numbers of \\ demonstrations in Self-Prompting.}
        \label{fig:vary-k}
    \end{minipage}%
    % Right graph
    \begin{minipage}{.445\textwidth}
        \centering
        \includegraphics[width=0.95\linewidth]{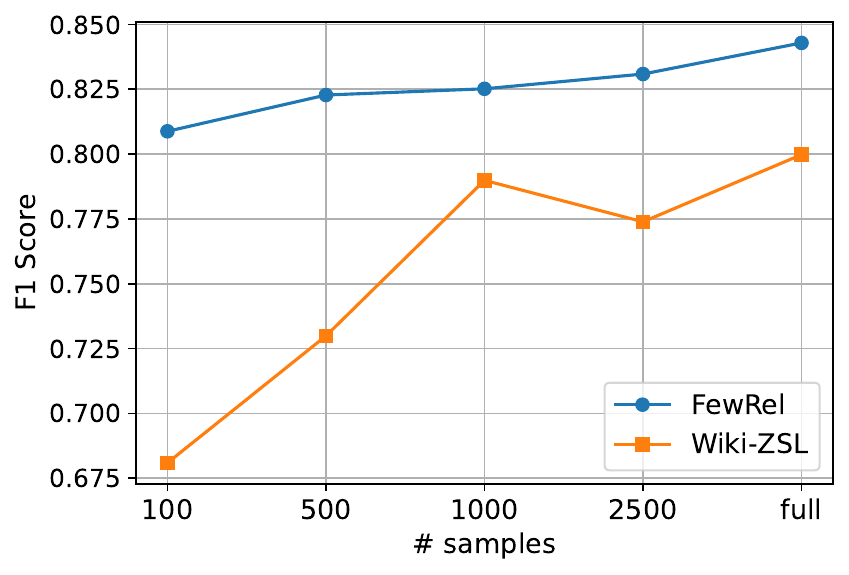}
        
        \caption{Average F1 when using different sizes of synthetic samples in Self-Prompting.}
        \label{fig:data-size}
    \end{minipage}
\end{figure*}

\section{Hyperparameter Settings}
\label{hp-setting}
During the synthetic sample generation phase, the temperature setting was adjusted to 1.2 to enhance sample diversity. Conversely, for inference, we set the temperature to 0, ensuring reproducibility, with other hyperparameters maintained at default settings.

For generating relation synonyms, we produced 10 synonyms per relation ($k = 10$). In the synthetic sample generation and filtering process, the LLMs were prompted to generate 10 samples at a time, excluding those with entities occurring more than three times ($n=3$). The generation process ceased either upon reaching 200 samples or when no new samples contained unique entities after three iterations for each relation. Each sample underwent sentence rephrasing to generate three variants ($r=3$). A detailed cost analysis is provided in Appendix \ref{sec:cost}. Regarding the selection of demonstration samples at inference, we fixed $d$ at 10. Following \citet{kojima2022large}, our approach only retains the first part of the model's output that conforms to the specified answer format.

\section{Influence of Demonstration Quantity}

To identify the optimal number of in-context samples $d$, we analyzed how varying the number of examples in the input affects performance, as illustrated in Figure \ref{fig:vary-k}. These experiments, aimed at assessing cost-effectiveness, were limited to a single subset of relations with $m = 10$. Analyzing F1 scores across two datasets revealed a pattern of performance improvement as the number of examples increased from 1 to 12. Yet, we found that utilizing more than 10 examples did not offer substantial benefits and, notably for Wiki-ZSL, resulted in diminished performance. Therefore, balancing performance efficiency with cost considerations, we determined that 10 demonstrations ($d=10$) were optimal for our experiments.

\section{Influence of Generated Data Size}

Evaluating the impact of synthetic sample size on experimental outcomes, our comprehensive analysis, shown in Figure \ref{fig:data-size}, focuses on a relation subset with $m=10$, exploring synthetic sample sizes from 100 to approximately 6,000.

The analysis reveals a clear trend: an increase in synthetic sample size generally boosts the F1 score across both FewRel and Wiki-ZSL datasets. Specifically, the FewRel dataset shows a steady increase in performance, reaching its peak with the full dataset utilized. In contrast, the Wiki-ZSL dataset experiences a marked improvement in F1 scores from 100 to 1,000 samples, after which the gains taper off, with scores stabilizing at 2,500 samples and beyond. This indicates that while enlarging the synthetic sample pool enhances model performance, a saturation point exists beyond which no significant benefits are observed.

\section{Data Generation Quality Analysis}
\begin{figure}[!ht]
    \centering
    \includegraphics[width=0.8\linewidth]{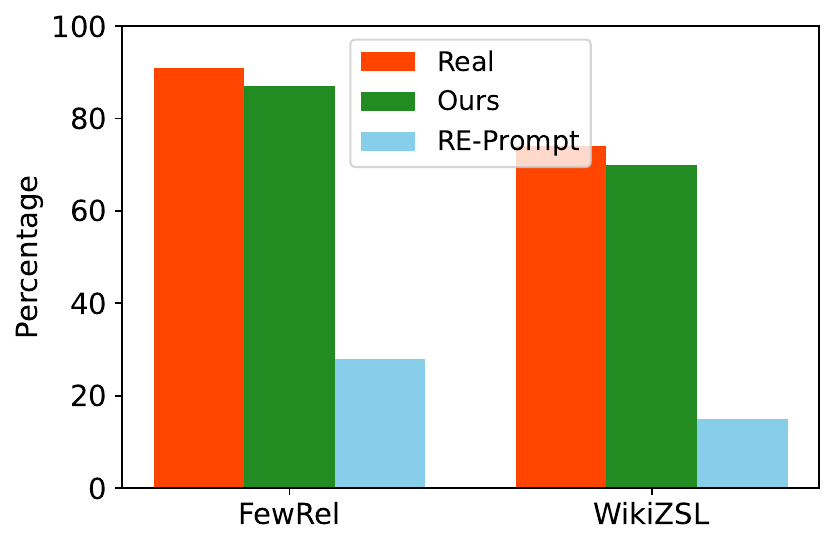}
    \caption{Percentage of correct samples in FewRel and Wiki-ZSL}
    \label{fig:quality}
\end{figure}
We employed GPT-4 to determine the presence of specified relations within various datasets to evaluate the quality of generated samples. We randomly selected 10 relations from each dataset, generating 10 samples for each, thereby creating a set of 100 samples per dataset. This analysis encompassed three datasets: the original real data, our generated data, and data generated using the RE-Prompt method. GPT-4 was tasked with verifying the specified relations in these samples. A sample was deemed correct if the head and tail entities exhibited the relation as labeled.

Figure \ref{fig:quality} shows that our generated samples more accurately encapsulate the targeted relations compared to those generated by the RE-Prompt method. This close alignment with real data benchmarks demonstrates the effectiveness of our generation methodology, validating our samples' utility for in-context learning in RE tasks.

\section{Comparing among Different Demonstration Data}
To further compare the quality of synthetic data from our method against RE-Prompt, we utilized RE-Prompt's synthetic data as demonstration samples in our inference framework. We documented the experimental outcomes on the FewRel and Wiki-ZSL datasets, with $m=10$, in Table \ref{table:compare}. These outcomes uniformly demonstrate that our method surpasses RE-Prompt in all instances, highlighting the superior data quality generated by our approach. This advantage is attained without task-specific fine-tuning, showcasing our data generation process's ability to produce high-quality synthetic samples for RE tasks effectively.
\begin{table}[h]
\centering

\resizebox{0.5\textwidth}{!}{
    \begin{tabular}{l ccc | ccc}
    \toprule
    Datasets & \multicolumn{3}{c}{ FewRel } & \multicolumn{3}{c}{ Wiki-ZSL } \\

    & Prec. & Rec. & F1 & Prec. & Rec. & F1 \\
    \midrule 
    Vanilla & 82.51 & 78.32 & 80.36 & 68.50 & 72.23 & 70.31 \\
    RE-Prompt & 83.73 & 81.30 & 82.50 & 73.33 & 72.14 & 72.73 \\
    Self-Prompting & \textbf{85.47} & \textbf{83.13} & \textbf{84.28} & \textbf{83.64} & \textbf{76.54} & \textbf{79.93} \\
    \bottomrule
    \end{tabular}}
\caption{Performance on FewRel and Wiki-ZSL datasets using varied synthetic demonstrations with \( m=10 \) unseen relations}
\label{table:compare}
\end{table}

\begin{table*}[h]
\centering
\resizebox{0.95\textwidth}{!}{
    \begin{tabular}{l l ccc | ccc | ccc | c}
    \toprule
    \multirow{2}{*}{Type} & \multirow{2}{*}{Method} & \multicolumn{3}{c}{$m=5$} & \multicolumn{3}{c}{$m=10$} & \multicolumn{3}{c}{$m=15$} & \multirow{2}{*}{Avg. Improv.}\\
 
    & & Prec. & Rec. & F1 & Prec. & Rec. & F1 & Prec. & Rec. & F1 \\
    \midrule
    \multirow{2}{*}{Qwen-1.8B} & Vanilla & 51.23 & 47.47 & 49.28 & 22.81 & 27.36 & 24.89 & 20.75 & 24.42 & 22.49 & \multirow{2}{*}{14.57\%} \\
    & Self-Prompting & 59.30 & 59.28 & 59.29 & 47.31 & 46.80 & 47.05 & 33.66 & 34.43 & 34.04  & \\
    \midrule
    
    \multirow{2}{*}{Qwen-7B} & Vanilla & 64.85 & 62.60 & 63.69 & 37.80 & 40.24 & 38.98 & 27.71 & 30.05 & 28.82 & \multirow{2}{*}{10.07\%}\\
    & Self-Prompting & 64.09 & 65.49 & 64.78 & 54.85 & 55.85 & 55.35 & 41.97 & 41.20 & 41.58 & \\
    \midrule

    \multirow{2}{*}{Qwen-14B} & Vanilla & 66.13 & 65.20 & 65.66 & 53.03 & 52.31 & 52.67 & 47.73 & 45.60 & 46.64 & \multirow{2}{*}{6.63\%}\\
    & Self-Prompting & 75.00 & 69.86 & 72.33 & 63.17 & 60.05& 61.67 & 51.70 & 50.03 & 50.85 & \\
    \midrule
    
    \multirow{2}{*}{ChatGPT} & Vanilla & 91.70 & 88.87 & 90.26 & 72.64 & 76.12 & 74.34 & 65.46 & 65.50 & 65.48 & \multirow{2}{*}{5.24\%}\\
    & Self-Prompting & 88.47 & 88.92 & 88.70 & 80.27 & 82.08 & 81.17 & 74.82 & 77.05 & 75.92 & \\
    
    \bottomrule
    \end{tabular}
}
\caption{Performance of our method for LLMs with different size}
\label{table:qwen}
\end{table*}

\section{Cost of Synthetic Data Generation}
\label{sec:cost}

For synthetic data generation, we employed \texttt{gpt-3.5-turbo}, an economical choice at \$0.001 per 1K tokens for prompts and \$0.002 per 1K tokens for completions\footnote{\url{https://openai.com/pricing}}. The synthesis involves three phases: generating relation synonyms, creating samples, and rephrasing sentences. The costs for each relation's data generation are itemized in Table \ref{table:cost}, totaling approximately \$0.264 for around 600 samples per relation. Considering the Wiki-ZSL dataset includes up to 113 relations, the full data generation cost is estimated at \$30. This is cost-effective compared to manual annotation expenses, such as in machine translation tasks, which can reach around \$0.1 per word \cite{Neubig_Zeno_GPT_Machine_2023}. Thus, using \texttt{gpt-3.5-turbo} for synthetic data generation in RE tasks is validated as an economically viable method.

\begin{table}[h]
    \centering
    \resizebox{0.5\textwidth}{!}{
        \begin{tabular}{lcccc}
        \toprule
        Stage & \# Prompt  & \# Completion & \# Total  & Cost (\$) \\
        \midrule
        Relation Synonyms & 0.132 & 0.077 & 0.209 & 0.00029 \\
        Sample Generation & 38.18 & 23.14 & 61.33 & 0.08447  \\
        Sentence Rephrase & 112.58 & 33.55 & 146.12 & 0.17967 \\
        \midrule
        Total & 150.89 & 56.77 & 207.66 & 0.26443 \\
        \bottomrule
    \end{tabular}}
\caption{Average number of token usage (k) and cost (\$) for a single relation samples generation}
\label{table:cost}
\end{table}

\section{General Effectiveness with LLMs of Different Sizes}
To examine the impact of LLM size, we also employed the Qwen \cite{bai2023qwen} series LLMs (1.8B, 7B, 14B) as alternative base models for evaluating our Self-Prompting methods. Our research explored Self-Prompting's efficacy across LLMs of various sizes, with the findings detailed in the accompanying table. This analysis covered models ranging from Qwen-1.8B to ChatGPT, applying both Vanilla and Self-Prompting methods to different sets of unseen relations (\(m = 5, 10, 15\)) in the FewRel dataset.

The Qwen series models (1.8B, 7B, and 14B parameters) demonstrated clear enhancements using Self-Prompting compared to the Vanilla approach. For the smallest model, Qwen-1.8B, Self-Prompting achieved a 14.57\% average increase in F1 scores, highlighting its significant benefit for smaller-scale models. With larger models, the average improvement lessened but remained impactful: 10.07\% for Qwen-7B and 6.63\% for Qwen-14B.

\section{Case Study}
\textbf{Generation:}
Tables \ref{table:case-study-1} and \ref{table:case-study-2} showcase examples of the generation process for the \textit{location} and \textit{operator} relations, respectively. \\
\textbf{Inference:} Table \ref{table:case-study-inf-1} presents a successful instance of Self-Prompting, while Table \ref{table:case-study-inf-2} illustrates a failure. The success case demonstrates how synthetic in-context samples, when closely related to the test sample, can offer a nuanced guide, aiding the model in distinguishing between \textit{location} and \textit{located on terrain feature}. Conversely, in the failure case, Self-Prompting did not yield an accurate prediction due to the in-context samples being less relevant, thereby introducing noise during inference.

\begin{table*}[h]
\centering
\resizebox{\textwidth}{!}{
    \begin{tabular}{p{0.1\textwidth} | p{0.85\textwidth}}
    % \hline
    \toprule
    Stage & Examples \\
    \midrule
    \multirow{4}{=}{Relation Synonyms} & \textbf{Relation}: \textcolor{cyan}{Location} \\
    & \textbf{Description}: location of the item, physical object, or event is within. \\
    & \textbf{Synonyms}: [situated at, found within, positioned in, nestled amongst, geographically placed, lying in, set within, residing at, located near, anchored in] \\
    \midrule
    \multirow{5}{=}{Sample Generation} & \textbf{Relation}: \textcolor{cyan}{Location} \\
    & 1. The \textcolor{orange}{grocery store} in my \textcolor{orange}{neighborhood} has a wide variety of organic produce. \\
    & 2. The \textcolor{orange}{rainforest}, filled with exotic wildlife, is set within the \textcolor{orange}{Amazon River basin}. \\
    & 3. The \textcolor{orange}{Louvre Museum}, one of the world's largest art museums, sits within the city of \textcolor{orange}{Paris}. \\
    \midrule
    \multirow{9}{=}{Rephrase Sentence} & \textbf{Relation}: \textcolor{cyan}{Location} \\
    & \textbf{Sentence}: The historic \textcolor{orange}{Colosseum} is set within the heart of \textcolor{orange}{Rome}, surrounded by ancient ruins and archaeological sites. \\
    & \textbf{Rephrased Sentence:} \\
    & 1. At the core of \textcolor{orange}{Rome}, the \textcolor{orange}{Colosseum} stands amidst ancient ruins and archaeological wonders.\\
    & 2. Surrounded by relics of the past, the \textcolor{orange}{Colosseum} exists at the center of  \textcolor{orange}{Rome}, a city with a rich history. \\
    & 3. \textcolor{orange}{Rome's} heart holds the majestic \textcolor{orange}{Colosseum}, encircled by remnants of the ancient era. \\
    \bottomrule
    \end{tabular}
}
\caption{Case of sample generation for relation \textbf{Location}}
\label{table:case-study-1}
\end{table*}

\begin{table*}[h]
\centering
\resizebox{\textwidth}{!}{
    \begin{tabular}{p{0.1\textwidth} | p{0.85\textwidth}}
    % \hline
    \toprule
    Stage & Examples \\
    \midrule
    \multirow{5}{=}{Relation Synonyms} & \textbf{Relation}: \textcolor{cyan}{Operator} \\
    & \textbf{Description}: person, profession, or organization that operates the equipment, facility, or service. \\
    & \textbf{Synonyms}: [controller, manager, handler, technician, operator, administrator, machinist, supervisor, system operator, service provider] \\
    \midrule
    \multirow{6}{=}{Sample Generation} & \textbf{Relation}: \textcolor{cyan}{Operator} \\
    & 1. The \textcolor{orange}{doctor}, who works at the hospital, is responsible for overseeing the \textcolor{orange}{medical equipment}. \\
    & 2. The \textcolor{orange}{IT technician} is in charge of maintaining and operating the \textcolor{orange}{computer server}. \\
    & 3. The \textcolor{orange}{internet connection} provided by the \textcolor{orange}{telecommunications company} has been unreliable lately. \\
    \midrule
    \multirow{6}{=}{Rephrase Sentence} & \textbf{Relation}: \textcolor{cyan}{Operator} \\
    & \textbf{Sentence}: The \textcolor{orange}{train station} is operated by the \textcolor{orange}{city transportation authority}. \\
    & \textbf{Rephrased Sentence:} \\
    & 1. The \textcolor{orange}{train station} falls under the jurisdiction of the \textcolor{orange}{city transportation authority}.\\
    & 2. The \textcolor{orange}{city transportation authority} oversees the operations of the \textcolor{orange}{train station}. \\
    & 3. The \textcolor{orange}{city transportation authority} is in charge of managing the \textcolor{orange}{train station}. \\
    \bottomrule
    \end{tabular}
}
\caption{Case of sample generation for relation \textbf{Operator}}
\label{table:case-study-2}
\end{table*}

% case: fewrel01 309
\begin{table*}[h]
\centering
\resizebox{\textwidth}{!}{
    \begin{tabular}{p{0.12\textwidth} | p{0.85\textwidth}}
    % \hline
    \toprule
    Stage & Examples \\
    \midrule
    \multirow{6}{=}{Background Prompts} & \textbf{Relation}: You are a helpful information extractor that can conduct relation extraction task. In detail, you final goal is to extract the relation between two entities in a sentence. The relation candidate is a list of  relations that you can choose from: \\
    & ['religion', 'location', 'competition class', 'operating system', 'owned by', 'contains administrative territorial entity', 'field of work', 'spouse', 'located on terrain feature', 'distributed by'] \\
    \midrule
    \multirow{16}{=}{Synthetic In-Context Prompts} 
    & \textbf{Sentence}: The \textcolor{orange}{ski resort town}, nestled against the natural feature of \textcolor{orange}{snow-capped mountains}, is a popular destination for winter sports enthusiasts. \\
    & Given the Sentence, the relation between \textcolor{orange}{town} and \textcolor{orange}{snow-capped mountains} is: \textcolor{cyan}{located on terrain feature} \\
    & \textbf{Sentence}: The \textcolor{orange}{village}, with its enchanting \textcolor{orange}{vineyards} and stunning vistas, finds itself nestled in the picturesque \textcolor{orange}{valley}. \\
    & Given the Sentence, the relation between \textcolor{orange}{village} and \textcolor{orange}{valley} is: \textcolor{cyan}{location} \\
    & \textbf{Sentence}: The beautiful \textcolor{orange}{vineyard}, with rolling hills as its backdrop, is situated near the quaint \textcolor{orange}{village} and nearby tourist destinations. \\
    & Given the Sentence, the relation between \textcolor{orange}{vineyard} and \textcolor{orange}{village} is: \textcolor{cyan}{location} \\
    & \textbf{Sentence}: Perched on the \textcolor{orange}{hill}, the \textcolor{orange}{building} provides a stunning vista of the \textcolor{orange}{valley} beneath. \\
    & Given the Sentence, the relation between \textcolor{orange}{building} and \textcolor{orange}{hill} is: \textcolor{cyan}{located on terrain feature} \\
    & \textbf{Sentence}: Renowned for its geysers and hot springs, \textcolor{orange}{Yellowstone National Park} is situated in the \textcolor{orange}{western United States}.  \\
    & Given the Sentence, the relation between \textcolor{orange}{Yellowstone National Park} and \textcolor{orange}{western United States} is: \textcolor{cyan}{located on terrain feature} \\

    \midrule
    \multirow{3}{=}{Test Sample Prompt} 
    & \textbf{Sentence}: It is located west of, and adjacent to \textcolor{orange}{Bridalveil Fall}, on the south side of the Merced River in \textcolor{orange}{Yosemite Valley}. \\
    & Given the Sentence, the relation between \textcolor{orange}{Bridalveil Fall} and \textcolor{orange}{Yosemite Valley} is: \\
    \midrule
    \multirow{3}{=}{Output} 
    & \textbf{Ground truth}: located on terrain feature  \\
    & \textbf{Vanilla}: location \textcolor{crosscolor}{\xmark}\\
    & \textbf{Self-Prompting}: located on terrain feature \textcolor{checkcolor}{\cmark}\\
    \bottomrule
    \end{tabular}
}
\caption{Case of successful test sample inference}
\label{table:case-study-inf-1}
\end{table*}

% case: fewrel01 207
\begin{table*}[h]
\centering
\resizebox{\textwidth}{!}{
    \begin{tabular}{p{0.12\textwidth} | p{0.85\textwidth}}
    % \hline
    \toprule
    Stage & Examples \\
    \midrule
    \multirow{6}{=}{Background Prompts} & \textbf{Relation}: You are a helpful information extractor that can conduct relation extraction task. In detail, you final goal is to extract the relation between two entities in a sentence. The relation candidate is a list of  relations that you can choose from: \\
    & ['religion', 'location', 'competition class', 'operating system', 'owned by', 'contains administrative territorial entity', 'field of work', 'spouse', 'located on terrain feature', 'distributed by'] \\
    \midrule
    \multirow{16}{=}{Synthetic In-Context Prompts} 
    & \textbf{Sentence}: An operating system known as \textcolor{orange}{macOS} powers the \textcolor{orange}{Mac computers}, which are produced by \textcolor{orange}{Apple Inc.} \\
    & Given the Sentence, the relation between \textcolor{orange}{computers} and \textcolor{orange}{Mac} is: \textcolor{cyan}{operating system}\\
    & \textbf{Sentence}: \textcolor{orange}{Linux}, a widely used \textcolor{orange}{open-source} operating system, is favored by programmers and developers. \\
    & Given the Sentence, the relation between \textcolor{orange}{Linux} and \textcolor{orange}{open-source} is: \textcolor{cyan}{operating system}\\
    & \textbf{Sentence}: The \textcolor{orange}{Unix operating system}, known for its stability and security, is widely used in \textcolor{orange}{enterprise computer systems}. \\
    & Given the Sentence, the relation between \textcolor{orange}{Unix operating system} and \textcolor{orange}{computer systems} is: \textcolor{cyan}{operating system}\\
    & \textbf{Sentence}: \textcolor{orange}{Windows}, commonly known as \textcolor{orange}{Microsoft Windows}, is a group of several proprietary graphical operating system families. \\
    & Given the Sentence, the relation between \textcolor{orange}{Windows} and \textcolor{orange}{Microsoft} is: \textcolor{cyan}{operating system}\\
    & \textbf{Sentence}: The construction and distribution of the iconic \textcolor{orange}{Lego sets} are handled by \textcolor{orange}{The Lego Group}, a Danish toy production company. \\
    & Given the Sentence, the relation between \textcolor{orange}{Lego sets} and \textcolor{orange}{The Lego Group} is: \textcolor{cyan}{distributed by}\\

    \midrule
    \multirow{4}{=}{Test Sample Prompt} 
    & \textbf{Sentence}: Sentence: His muscle algorithms for face animation were widely used in the computer film industry, most notably by \textcolor{orange}{Pixar}, which first used the technique in their animation short \textcolor{orange}{Tin Toy}. \\
    & Given the Sentence, the relation between \textcolor{orange}{Tin Toy} and \textcolor{orange}{Pixar} is: \\
    \midrule
    \multirow{3}{=}{Output} 
    & \textbf{Ground truth}: distributed by  \\
    & \textbf{Vanilla}: distributed by \textcolor{checkcolor}{\cmark}\\
    & \textbf{Self-Prompting}: field of work \textcolor{crosscolor}{\xmark}\\
    \bottomrule
    \end{tabular}
}
\caption{Case of failed test sample inference}
\label{table:case-study-inf-2}
\end{table*}

\section{Prompts for LLMs}

We listed each stage's prompts used in the synthetic data generation process in Table \ref{table:prompt}.
\begin{table*}[h]
\centering
\resizebox{\textwidth}{!}{
    \begin{tabular}{p{0.15\textwidth} | p{0.8\textwidth}}
    % \hline
    \toprule
    Stage & Prompts \\
    \midrule
    \multirow{9}{=}{Relation Synonyms} & For a giving relation type: \{\textit{relation}\}, your objective is to create \{\textit{k}\} synonyms about this relation. \\
    & The description of this relation is: \{\textit{description}\} \\
    & Ensure that your generated examples adhere to the following guidelines: \\
    & 1. The synonyms should explicitly or implicitly align with the relation \{\textit{relation}\}. \\
    & 2. Ensure the diversity among different synonyms. \\
    & 3. The synonyms could be a single word or phrase. \\
    & Please format your output in Python list-style: \\
    & [synonyms1, synonyms2, ..., synonyms\{\textit{k}\}] \\
    \midrule
    \multirow{17}{=}{Sample Generation} & Imagine you are a sophisticated language model functioning as a textual data generator for a relation extraction task. Your objective is to create \{\textit{k}\} synthetic sentences, each containing a specific type of relationship denoted as: \{\textit{relation}\} \\
    & The description of this relation is: \{\textit{description}\}. \\
    & These sentences must be informative and clearly demonstrate the intended relation, either explicitly or implicitly. Please format your output as follows: \\
    & Sentence: [Your generated sentence here]. \\
    & Relation: [(entity1, \{\textit{relation}\}, entity2), (entity3, \{\textit{relation}\}, entity4), ...]. \\
    & Where the relation list could contain one to three relation tuples. \\
    & Ensure that your generated examples adhere to the following guidelines: \\
    & 1. The relation should be the same as the previously defined relation. \\
    & 2. Head and tail entities must appear in the original sentence. \\
    & 3. Separate the head and tail into several triples that have the same relation. \\
    & 4. Generate sentences with varying lengths and complexities, including simple, compound, and complex sentences. \\
    & 5. Ensure a broad and realistic variety in the types of head and tail entities to reflect real-world contexts. \\
    \midrule
    \multirow{18}{=}{Rephrase Sentence} & As a text paraphrasing agent, your task is to paraphrase a given sentence to generate \{\textit{k}\} new versions. The original sentence includes one or more relationships. Rewrite the sentence to subtly imply the relationships that were originally stated explicitly, while also enhancing the semantic depth and diversifying the grammatical structure. \\
    & Input format: \\
    & Sentence: The sentence to be paraphrased. \\
    & Relation: A list of relation tuples in the format (head, relation, tail). \\
    & Output Format: \\
    & Provide \{\textit{k}\} paraphrased sentences, where the relation list could contain one to three relation tuples. \\
    & Ensure that your generated examples adhere to the following guidelines: \\
    & 1. Preservation of Entities: Ensure that the head and tail entities from the original sentence are present in each paraphrased version. \\
    & 2. Variety and Realism: Aim for a wide range of sentence structures and contexts in your paraphrases, reflecting realistic and diverse scenarios. \\
    & 3. In the generated relation list for each paraphrased sentence, the relation MUST remain consistent with the relation: \{\textit{relation}\}, while minor modifications to the entities are permissible. \\
    \midrule
    \multirow{5}{=}{Inference} & Your goal is to extract the relation between two entities in a sentence. The relation candidate is a list of relations that you can choose from: \{\textit{relation list}\} \\
    & \{\textit{demonstrations}\} \\
    & Sentence: \{\textit{extract sentence}\} \\
    & Given the Sentence, the relation between \{\textit{head}\} and \{\textit{tail}\} is: \\
    \bottomrule
    \end{tabular}
}
\caption{Prompts used for synthetic data generation and test sample inference}
\label{table:prompt}
\end{table*}

\end{document}